\documentclass[a4paper,10pt,eqno]{article}
\usepackage{theorem}
\usepackage{latexsym,amssymb,amsfonts,amsmath}
\usepackage{cite}
\newtheorem{thm}{Theorem}[section]
\newtheorem{lem}[thm]{Lemma}

\newtheorem{pro}[thm]{Proposition}
\newtheorem{rem}[thm]{Remark}

\newtheorem{Def}[thm]{Definition}

\theoremheaderfont{\scshape}
\newcommand{\RM}{\mathbb{R}}

\newcommand{\PM}{\mathbb{P}}

\newcommand{\qed}{\hfill $\Box$}

\title{{\Large {\bf Local subgraph structure can cause localization in continuous-time quantum walk}}}
\author{
{\small Yusuke Ide}\\
{\scriptsize Department of Information Systems Creation, 
Faculty of Engineering, 
Kanagawa University}\\
{\scriptsize Kanagawa, Yokohama 221-8686, Japan}\\
{\scriptsize e-mail: ide@kanagawa-u.ac.jp}
}
\date{\empty }
\begin{document}
\maketitle

\par\noindent
\begin{small}
\par\noindent
{\bf Abstract}
\newline 
In this paper, we consider continuous-time quantum walks (CTQWs) on finite graphs determined by the Laplacian matrices. By introducing fully interconnected graph decomposition of given graphs, we show a decomposition method for the Laplacian matrices. Using the decomposition method, we show several conditions for graph structure which return probability of CTQW tends to 1 while the number of vertices tends to infinity.
\end{small}

\setcounter{equation}{0}

\section{Introduction}
Quantum walks (QWs) have been attractive research topic in this decade \cite{Konno2008b, VAndraca2012, ManouchehriWang2013} as quantum counterparts of the random walks which play important roles in various fields. For QWs, there are two types of time evolution, discrete-time and continuous-time. In this paper, we focus on continuous-time quantum walks (CTQWs) on finite graphs. There are a lot of studies of CTQWs on various deterministic graphs, such as the line \cite{Konno2005}, path graph \cite{Godsil2013}, star graph \cite{Salimi2009, Xu2009}, cycle graph \cite{abtw2003,ikkk2005,MuelkenBlumen2006}, dendrimers \cite{mbb2006}, spidernet graphs \cite{Salimi2010QIP}, the dual Sierpinski gasket \cite{abm2008}, direct product of Cayley graphs \cite{SalimiJafarizadeh2009}, quotient graphs \cite{Salimi2008IJQI}, odd graphs \cite{Salimi2008IJTP}, trees \cite{Konno2006IDAQP,JafarizadehSalimi2007} and ultrametric spaces \cite{Konno2006IJQI}. Also there are studies of CTQWs on probabilistic graphs, such as small-world networks \cite{mpb2007}, Erd\H{o}s-R\'enyi random graph \cite{XuLiu2008} and the threshold network model \cite{IdeKonno2010, KirklandSeverini2011}. 

Here we give the definition of our CTQW.  Let $G_{n}$ be a simple (undirected) graph with $n$ numbers of vertices. In this paper, we use $V(G_{n})=\{1,\ldots,n\}$ for the vertex set and $E(G_{n})\subset V(G_{n})\times V(G_{n})$ for the edge set of the graph $G_{n}$. For a pair of vertices $i,j\in V(G_{n})$, we write $i\sim j$ if $(i,j)\in E(G_{n})$, i.e., the pair of vertices $i$ and $j$ is connected by an edge. Let $A_{G_{n}}$ be the adjacency matrix of the graph $G_{n}$ which is an $n\times n$ matrix whose $(i,j)$ component $(A_{G_{n}})_{i,j}$ equals $1$ if $i\sim j$ and $0$ otherwise. The Laplacian matrix $L_{G_{n}}$ of $G_{n}$ is defined by $L_{G_{n}}=D_{G_{n}}-A_{G_{n}}$ where $D_{G_{n}}$ be the $n\times n$ diagonal matrix given by $D_{G_{n}}=\mathrm{diag}(d_{G_{n}}(1),\ldots ,d_{G_{n}}(n))$ with $d_{G_{n}}(i)=\sum _{j=1}^{n}(A_{G_{n}})_{i,j}$, i.e., the degree of the vertex $i$, for $i\in V(G_{n})$.

The time evolution operator $U_{G_{n},t}$ of a CTQW on $G_{n}$ at time $t\geq 0$ is defined by 
\begin{eqnarray}\label{defU}
U_{G_{n},t}\equiv e^{\sqrt{-1}tL_{G_{n}}}= \sum _{k=0}^{\infty }\frac{(\sqrt{-1}t)^{k}}{k!}L_{G_{n}}^{k},
\end{eqnarray}
where $\sqrt{-1}$ be the imaginary unit. Let $\{\Psi _{G_{n},t}\}_{t\geq 0}$ be the probability amplitude of the quantum walk, i.e.,
$
\Psi _{G_{n},t}=U_{G_{n},t}\Psi _{G_{n},0},
$
where 
$
\Psi _{G_{n},0}=
{}^{T}\!
\begin{bmatrix}
\Psi _{G_{n},0}(1) & \ldots & \Psi _{G_{n},0}(n)
\end{bmatrix}
$
is an $n$ dimensional unit vector which we call the initial condition where ${}^TA$ is the transpose of a matrix $A$. Then the probability that the quantum walker on $G_{n}$ is in position $y\in V(G_{n})$ 
at time $t$ with initial condition $\Psi _{G_{n},0}$ is defined by 
\begin{eqnarray*}
\PM(Y_{G_{n},t}^{\Psi _{G_{n},0}}=y)\equiv |(U_{G_{n},t}\Psi _{G_{n},0})(y)|^{2},
\end{eqnarray*}
where $Y_{G_{n},t}^{\Psi _{G_{n},0}}$ be the random variable representing the quantum walker' s position at time $t$ on $G_{n}$ with initial condition ${\Psi _{G_{n},0}}$. In this paper, we only deal with $\Psi _{G_{n},0}(x)=1$ for some specific vertex $x\in V(G_{n})$ and $\Psi _{G_{n},0}(x^{\prime })=0$ for $x^{\prime} \neq x$ case. Note that this corresponds to the case that the walker starts from the vertex $x$.  Hereafter, we use $P_{G_{n},t}^{x}(y)$ instead of $\PM(Y_{G_{n},t}^{\Psi _{G_{n},0}}=y)$ for simplicity. 

In this paper, we call strong localization for $x\in V(G_{n})$ occur when the return probability tends to $1$ in $n\to \infty$, i.e., 
\begin{align*}
\lim_{n\to \infty }P_{G_{n},t}^{x}(x)=1.
\end{align*}
It is known that CTQWs defined by the Laplacian matrix on complete graphs (see e.g. \cite{Konno2008b}), star graphs\cite{Salimi2009, Xu2009} and the threshold network model \cite{IdeKonno2010, KirklandSeverini2011} have the same transition probabilities from the vertices which connect with all other vertices and also strong localization for the vertices occur. But it seems that there are no comprehensive treatments for relationships between graph structure and the transition probabilities of such graphs. 

The aim of this paper is to clarify relationships between graph structure and the transition probabilities of CTQWs on graphs. In order to do so, we introduce fully interconnected graph decomposition (Definition \ref{defFID}) which is a generalization of the graph operation ``join'' in Sec.\ 2. We should note that the decomposition procedure for the Laplacian matrix proposed in Sec.\ 2 is motivated by Merris' s work \cite{Merris1994, Merris1998}. After that we derive a decomposition formula for transition probabilities of CTQW with related to the decomposition (Lemma \ref{lem:transition}). As a consequence, we find that the limit of the return probabilities of the CTQW on graph with the decomposition starting from a vertex in a growing subgraph are equal to that of the subgraph (Theorem \ref{thm:localization}). This means that local subgraph structure can cause localization in the whole graph. We show two concrete examples of CTQWs which cause strong localization for some vertices in Sec.\ 3. The first one (Sec. 3.1) includes complete graphs, star graphs and the threshold network model cases. The second one (Sec. 3.2) shows that growing clique can cause the strong localization. It can be interesting future problems that to find necessary and sufficient condition of graph structure for the strong localization and to build a discrete-time version of this decomposition method.

\section{Fully interconnected graph decomposition}

In this paper, we consider the following decomposition of the graph $G_{n}$:
\begin{Def}[Fully interconnected graph decomposition]\label{defFID}
Let $G_{n}$ be a simple graph. Then $(G_{n_{1}},\ldots G_{n_{k}})$ is said to be a fully interconnected graph decomposition of $G_{n}$ if it satisfies the following conditions:
\begin{enumerate}
\item
Each $G_{n_{i}}$ is an induced subgraph of $G_{n}$, i.e., if $v\sim w$ in $G_{n}$ then $v\sim w$ in $G_{n_{i}}$ for all $v,w\in V(G_{n_{i}})\subset V(G_{n})$, on $n_{i}$ numbers of vertices for $i=1,\ldots k$. 
\item
$V(G_{n})=V(G_{n_{1}})\cup \cdots \cup V(G_{n_{k}})$ and $V(G_{n_{i}})\cap V(G_{n_{j}})=\emptyset $ for $i\neq j$.
\item
For each pair of subgraphs $(G_{n_{i}},G_{n_{j}})$ for $i\neq j$, one of the following conditions is hold:
\begin{enumerate}
\item
All pairs of vertices $(v,w)\in V(G_{n_{i}})\times V(G_{n_{j}})$ are connected. In this case, we call the pair of subgraphs $(G_{n_{i}},G_{n_{j}})$ is fully interconnected and represent $G_{n_{i}}\sim G_{n_{j}}$. 
\item
All pairs of vertices $(v,w)\in V(G_{n_{i}})\times V(G_{n_{j}})$ are disconnected. In this case, we call the pair of subgraphs $(G_{n_{i}},G_{n_{j}})$ is fully interdisconnected and represent $G_{n_{i}}\nsim G_{n_{j}}$.
\end{enumerate}
\end{enumerate}
\end{Def}
Remark that $(G_{n})$ is a trivial fully interconnected graph decomposition of $G_{n}$.

Now we consider a ($k$ blocks $\times$ $k$ blocks) block matrix $\widetilde{L}_{G_{n}}$ of $G_{n}$ with a fully interconnected graph decomposition $(G_{n_{1}},\ldots G_{n_{k}})$ defined as follows:
\begin{align}\label{eqblock}
(\widetilde{L}_{G_{n}})_{i,j\ \mathrm{block}}=
\begin{cases}
\widetilde{d}_{i}I_{n_{i}}, &\text{if $i=j$},\\
-J_{n_{i},n_{j}}, &\text{if $G_{n_{i}}\sim G_{n_{j}}$},\\
O_{n_{i},n_{j}}, &\text{otherwise},
\end{cases}
\end{align}
where $\widetilde{d}_{i}=\sum _{G_{n_{i}}\sim G_{n_{j}}}n_{j}$, $I_{n}$ is the $n\times n$ identity matrix, $J_{l,m}$ is $l\times m$ all $1$ matrix and $O_{l,m}$ is $l\times m$ all $0$ matrix. The Laplacian matrix $L_{G_{n}}$ of $G_{n}$ with related to a fully interconnected graph decomposition $(G_{n_{1}},\ldots G_{n_{k}})$ is decomposed into two ($k$ blocks $\times$ $k$ blocks) block matrices as follows:
\begin{align}\label{eqdecompose}
L_{G_{n}}=\mathrm{diag}(L_{G_{n_{1}}},\ldots ,L_{G_{n_{k}}})+\widetilde{L}_{G_{n}}.
\end{align}

In order to analyze the time evolution operator of CTQW, we discuss about the eigenspace of $L_{G_{n}}$. Let $\{\lambda_{i,l_{i}}\}_{l_{i}=1,\ldots ,n_{i}-1}$ be the eigenvalues of $L_{G_{n_{i}}}$ except for the trivial eigenvalue $0$ corresponding to $n_{i}$ dimensional all $1$ vector $\mathbf{1}_{n_{i}}$ for $i=1,\ldots ,k$. The corresponding eigenvectors $\{\mathbf{v}_{i,l_{i}}\}_{l_{i}=1,\ldots ,n_{i}-1}$ can be $n_{i}$ dimensional real unit vectors and orthogonal to each other and orthogonal to $\mathbf{1}_{n_{i}}$ since each  $L_{G_{n_{i}}}$ is an real symmetric matrix. By Eqs.\ (\ref{eqblock}), (\ref{eqdecompose}), if we define
\begin{align*}
\mathbf{w}_{i,l_{i}}={}^{T}\!
[
\overbrace{0,\ldots ,0}^{n_{1}+\cdots +n_{i-1}}, \mathbf{v}_{i,l_{i}}(1), \ldots , \mathbf{v}_{i,l_{i}}(n_{i}), \overbrace{0,\ldots ,0}^{n_{i+1}+\cdots +n_{k}}
],\quad \text{($l_{i}=1,\ldots ,n_{i}-1$)},
\end{align*}
for $i=1,\ldots ,k$, where $\mathbf{v}_{i,l_{i}}(j)$ denotes the $j$-th component of $\mathbf{v}_{i,l_{i}}$, then it is easy to see that
\begin{align*}
L_{G_{n}}\mathbf{w}_{i,l_{i}}=\left(\mathrm{diag}(L_{G_{n_{1}}},\ldots ,L_{G_{n_{k}}})+\widetilde{L}_{G_{n}}\right)\mathbf{w}_{i,l_{i}}=(\lambda _{i,l_{i}}+\widetilde{d}_{i})\mathbf{w}_{i,l_{i}}.
\end{align*}
Thus we have $(n-k)$ numbers of eigenvalues and corresponding orthonormal eigenvectors of $L_{G_{n}}$ from the Laplacian matrices $L_{G_{n_{i}}}\ (i=1,\ldots ,k)$ of subgraphs $G_{n_{1}},\ldots ,G_{n_{k}}$.

The remaining  $k$ numbers of eigenvectors are corresponding to all $1$ vectors $\mathbf{1}_{n_{1}},\ldots, \mathbf{1}_{n_{k}}$. Let 
\begin{align}\label{eqconstvec}
\mathbf{x}_{i}={}^{T}\!
[
\overbrace{\alpha_{i}(1),\ldots ,\alpha_{i}(1)}^{n_{1}}, \overbrace{\alpha_{i}(2),\ldots ,\alpha_{i}(2)}^{n_{2}}, \ldots , \overbrace{\alpha_{i}(k),\ldots ,\alpha_{i}(k)}^{n_{k}}
],
\end{align}
for $i=1,\ldots ,k$, where $\alpha _{i}(1),\ldots ,\alpha _{i}(k)\in \RM$. Then we have
\begin{align*}
L_{G_{n}}\mathbf{x}_{i}=\left(\mathrm{diag}(L_{G_{n_{1}}},\ldots ,L_{G_{n_{k}}})+\widetilde{L}_{G_{n}}\right)\mathbf{x}_{i}=\widetilde{L}_{G_{n}}\mathbf{x}_{i}.
\end{align*}
Note that from Eqs.\ (\ref{eqblock}), (\ref{eqconstvec}), the eigen equations $\widetilde{L}_{G_{n}}\mathbf{x}_{i}=\nu _{i}\mathbf{x}_{i}$ are equivalent to $\overline{L}_{G_{n}}\overline{\mathbf{x}}_{i}=\nu _{i}\overline{\mathbf{x}}_{i}$ with a $k\times k$ matrix $\overline{L}_{G_{n}}$ such that
\begin{align*}
\left(\overline{L}_{G_{n}}\right)_{i,j}=
\begin{cases}
\widetilde{d}_{i}, &\text{if $i=j$},\\
-n_{j}, &\text{if $G_{n_{i}}\sim G_{n_{j}}$},\\
0, &\text{otherwise},
\end{cases}
\end{align*}
and a $k$-dimensional vector
\begin{align*}
\overline{\mathbf{x}}_{i}={}^{T}\!
[
\alpha _{i}(1),\ldots ,\alpha _{i}(k)
].
\end{align*}

Because we can take the set of eigenvectors as an orthonormal base,  the following matrix $B_{G_{n}}$ can be an orthogonal matrix:
\begin{align*}
B_{G_{n}}\equiv 
\left[
\mathbf{w}_{1,1},\ldots ,\mathbf{w}_{1,n_{1}-1}, \ldots ,\mathbf{w}_{k,1},\ldots ,\mathbf{w}_{k,n_{k}-1}, \frac{\mathbf{x}_{1}}{\sqrt{\sum _{l=1}^{k}n_{l}\alpha _{1}(l)^{2}}},\ldots ,\frac{\mathbf{x}_{k}}{\sqrt{\sum _{l=1}^{k}n_{l}\alpha _{k}(l)^{2}}}
\right].
\end{align*}
After diagonalization of the time evolution operator $U_{G_{n},t}$ of CTQW on $G_{n}$ (Eq.\ (\ref{defU})) by using $B_{G_{n}}$, we have the following spectoral decomposition of $U_{G_{n},t}$:
\begin{align*}
&\left(U_{G_{n},t}\right)_{x,y}\notag\\
&=
\begin{cases}
\displaystyle\sum _{j=1}^{n_{i}-1}\exp\left\{\sqrt{-1}t(\lambda _{i,j}
+
\widetilde{d}_{i})\right\}\mathbf{v}_{i,j}(x)\mathbf{v}_{i,j}(y)\\
+
\displaystyle\sum _{j=1}^{k}\exp\left(\sqrt{-1}t\nu _{j}\right)\frac{\alpha _{j}(i)^{2}}{\sum _{l=1}^{k}n_{l}\alpha _{j}(l)^{2}}
&\text{if $x,y\in V(G_{n_{i}})$},
\\
\\
\displaystyle\sum _{j=1}^{k}\exp\left(\sqrt{-1}t\nu _{j}\right)\frac{\alpha _{j}(i)\alpha _{j}(i^{\prime })}{\sum _{l=1}^{k}n_{l}\alpha _{j}(l)^{2}}
&\text{if $x\in V(G_{n_{i}})$ and $y\in V(G_{n_{i^{\prime }}})$ ($i\neq i^{\prime }$)}.
\end{cases}
\end{align*}
Therefore we have the transition probabilities of CTQW as follows:
\begin{lem}\label{lem:transition}
Let $(G_{n_{1}}, \ldots ,G_{n_{k}})$ be a fully interconnected graph decomposition of a graph $G_{n}$. Then the transition probabilities of CTQW are given as follows:
\begin{align}\label{expressU2}
&P_{G_{n},t}^{x}(y)\notag\\
&=
\left|\left(U_{G_{n},t}\right)_{x,y}\right|^{2}\notag\\
&=
\begin{cases}
P_{G_{n_{i}},t}^{x}(y)
+
\widetilde{P}_{G_{n_{i}},t}^{x}(y)\\
-
\displaystyle\frac{1}{n_{i}^{2}}
-
\frac{2}{n_{i}}\sum _{j=1}^{n_{i}-1}\mathbf{v}_{i,j}(x)\mathbf{v}_{i,j}(y)\cos(t\lambda _{i,j})\\
+
2\displaystyle\sum _{j=1}^{n_{i}-1}\sum _{j^{\prime }=1}^{k}\frac{\mathbf{v}_{i,j}(x)\mathbf{v}_{i,j}(y)\alpha _{j^{\prime }}(i)^{2}\cos\left\{t(\lambda _{i,j}+\widetilde{d}_{i}-\nu _{j^{\prime }})\right\}}{\sum _{l=1}^{k}n_{l}\alpha _{j^{\prime }}(l)^{2}}
&\text{if $x,y\in V(G_{n_{i}})$},
\\
\\
\widetilde{P}_{G_{n_{i}},G_{n_{i^{\prime }}},t}^{x}(y)
&\text{if $x\in V(G_{n_{i}})$ and $y\in V(G_{n_{i^{\prime }}})$ ($i\neq i^{\prime }$)}.
\end{cases}
\end{align}
where 
\begin{align}
P_{G_{n_{i}},t}^{x}(y)
&=
\displaystyle\sum _{j=1}^{n_{i}-1}\mathbf{v}_{i,j}(x)^{2}\mathbf{v}_{i,j}(y)^{2}
+
\frac{1}{n_{i}^{2}}\notag\\
&+
2\displaystyle\sum _{1\leq j<j^{\prime }\leq n_{i}-1}\mathbf{v}_{i,j}(x)\mathbf{v}_{i,j}(y)\mathbf{v}_{i,j^{\prime }}(x)\mathbf{v}_{i,j^{\prime }}(y)\cos\left\{t(\lambda _{i,j}-\lambda _{i,j^{\prime }})\right\}\notag\\
&+
\displaystyle\frac{2}{n_{i}}\displaystyle\sum _{j=1}^{n_{i}-1}\mathbf{v}_{i,j}(x)\mathbf{v}_{i,j}(y)\cos(t\lambda _{i,j}),
\notag\\
\widetilde{P}_{G_{n_{i}},t}^{x}(y)
&=
\displaystyle\sum _{j=1}^{k}\frac{\alpha _{j}(i)^{4}}{\left(\sum _{l=1}^{k}n_{l}\alpha _{j}(l)^{2}\right)^{2}}
+
2\displaystyle\sum _{1\leq j<j^{\prime }\leq k}\frac{\alpha _{j}(i)^{2}\alpha _{j^{\prime }}(i)^{2}\cos\left\{t(\nu _{j}-\nu _{j^{\prime }})\right\}}{\left(\sum _{l=1}^{k}n_{l}\alpha _{j}(l)^{2}\right)\left(\sum _{l=1}^{k}n_{l}\alpha _{j^{\prime }}(l)^{2}\right)},
\label{defptildeni}\\
\widetilde{P}_{G_{n_{i}},G_{n_{i^{\prime }}},t}^{x}(y)
&=
\displaystyle\sum _{j=1}^{k}\frac{\alpha _{j}(i)^{2}\alpha _{j}(i^{\prime })^{2}}{\left(\sum _{l=1}^{k}n_{l}\alpha _{j}(l)^{2}\right)^{2}}
+
2\displaystyle\sum _{1\leq j<j^{\prime }\leq k}\frac{\alpha _{j}(i)\alpha _{j}(i^{\prime })\alpha _{j^{\prime }}(i)\alpha _{j^{\prime }}(i^{\prime })\cos\left\{t(\nu _{j}-\nu _{j^{\prime }})\right\}}{\left(\sum _{l=1}^{k}n_{l}\alpha _{j}(l)^{2}\right)\left(\sum _{l=1}^{k}n_{l}\alpha _{j^{\prime }}(l)^{2}\right)}.
\notag
\end{align}
\end{lem}
The first term $P_{G_{n_{i}},t}^{x}(y)$ in Eq.\ (\ref{expressU2}) is the transition probability of CTQW on the graph $G_{n_{i}}$. The second term $\widetilde{P}_{G_{n_{i}},t}^{x}(y)$ and the last term $\widetilde{P}_{G_{n_{i}},G_{n_{i^{\prime }}},t}^{x}(y)$ are the transition probabilities determined only by $\widetilde{L}_{G_{n}}$ which is not depend on the detailed structures of the subgraphs $G_{n_{1}}, \ldots ,G_{n_{k}}$. Because the number of vertices $n_{i}$ in $G_{n_{i}}$ plays an important role in Theorem \ref{thm:localization}, we explicitly describe $n_{i}$ in Eqs.\ (\ref{expressU2}) and  (\ref{defptildeni}). The following theorem shows that the terms in Eq.\ (\ref{expressU2}) except for $P_{G_{n_{i}},t}^{x}(y)$ vanish in $n_{i}\to \infty $ for return probability cases $(x=y)$:
\begin{thm}\label{thm:localization}
Let $(G_{n_{1}}, \ldots ,G_{n_{k}})$ be a fully interconnected graph decomposition of a graph $G_{n}$. If $\lim_{n_{i}\to \infty }P_{G_{n_{i}},t}^{x}(x)$ with $x\in V(G_{n_{i}})$ exists then
\begin{align*}
\lim_{n_{i}\to \infty }P_{G_{n,t}}^{x}(x)=\lim_{n_{i}\to \infty }P_{G_{n_{i}},t}^{x}(x).
\end{align*}
\end{thm}
(Proof of Theorem \ref{thm:localization})
\\
From $B_{G_{n}}\!{}^{T}\!B_{G_{n}}=I_{n}$, we have
\begin{align}\label{eqorth1}
\sum _{j=1}^{n_{i}-1}\mathbf{v}_{i,j}(x)\mathbf{v}_{i,j}(y)
+
\sum _{j=1}^{k}
\frac{\alpha _{j}(i)^{2}}{\sum _{l=1}^{k}n_{l}\alpha _{j}(l)^{2}}
=
\begin{cases}
1 &\text{if $x=y$,}\\
0 &\text{otherwise,}
\end{cases}
\end{align}
for $x,y\in V(G_{n_{i}})$, and also we have
\begin{align}\label{eqorth2}
\sum _{j=1}^{n_{i}-1}\mathbf{v}_{i,j}(x)\mathbf{v}_{i,j}(y)
+
\frac{1}{n_{i}}
=
\begin{cases}
1 &\text{if $x=y$,}\\
0 &\text{otherwise,}
\end{cases}
\end{align}
for $x,y\in V(G_{n_{i}})$ because $\{\mathbf{v}_{i,l_{i}}\}_{l_{i}=1,\ldots ,n_{i}-1}\cup \{\frac{1}{\sqrt{n_{i}}}\mathbf{1}_{n_{i}}\}$ is a set of orthonormal eigenvectors of $L_{G_{n_{i}}}$. Combining Eqs.\ (\ref{eqorth1}) and (\ref{eqorth2}), we obtain 
\begin{align}\label{eqorth3}
\sum _{j=1}^{k}
\frac{\alpha _{j}(i)^{2}}{\sum _{l=1}^{k}n_{l}\alpha _{j}(l)^{2}}
=\frac{1}{n_{i}}.
\end{align}
In particular, we obtain the following uniform bound from Eq.\ (\ref{eqorth3}):
\begin{align}\label{eqorth4}
\frac{\alpha _{j}(i)^{2}}{\sum _{l=1}^{k}n_{l}\alpha _{j}(l)^{2}}
\leq \frac{1}{n_{i}}\quad \text{for $\forall i,j\in \{1,\dots k\}$}.
\end{align}
By substituting Eq.\ (\ref{eqorth3}) into Eq.\ ({\ref{defptildeni}}) and using Eq.\ (\ref{eqorth4}), we have 
\begin{align*}
\widetilde{P}_{G_{n_{i}},t}^{x}(y)
&=
\displaystyle\sum _{j=1}^{k}\frac{\alpha _{j}(i)^{4}}{\left(\sum _{l=1}^{k}n_{l}\alpha _{j}(l)^{2}\right)^{2}}
+
2\displaystyle\sum _{1\leq j<j^{\prime }\leq k}\frac{\alpha _{j}(i)^{2}\alpha _{j^{\prime }}(i)^{2}\cos\left\{t(\nu _{j}-\nu _{j^{\prime }})\right\}}{\left(\sum _{l=1}^{k}n_{l}\alpha _{j}(l)^{2}\right)\left(\sum _{l=1}^{k}n_{l}\alpha _{j^{\prime }}(l)^{2}\right)}\notag\\
&\leq
\frac{1}{n_{i}}\displaystyle\sum _{j=1}^{k}\frac{\alpha _{j}(i)^{2}}{\sum _{l=1}^{k}n_{l}\alpha _{j}(l)^{2}}
+
2\left(\displaystyle\sum _{j=1}^{k}\frac{\alpha _{j}(i)^{2}}{\sum _{l=1}^{k}n_{l}\alpha _{j}(l)^{2}}\right)^{2}
=
\frac{3}{n_{i}^{2}}.
\end{align*}
%
%
%
%
%
%
On the other hand, by using Eqs.\ (\ref{eqorth1}) and (\ref{eqorth3}), we have
\begin{align*}
\frac{2}{n_{i}}\sum _{j=1}^{n_{i}-1}\mathbf{v}_{i,j}(x)^{2}\cos(t\lambda _{i,j})
&\leq 
\frac{2}{n_{i}}\left(1-\displaystyle\sum _{j=1}^{k}\frac{\alpha _{j}(i)^{2}}{\sum _{l=1}^{k}n_{l}\alpha _{j}(l)^{2}}\right)
=\frac{2}{n_{i}}\left(1-\frac{1}{n_{i}}\right),
\label{estpn1}
\\
2\displaystyle\sum _{j=1}^{n_{i}-1}\sum _{j^{\prime }=1}^{k}\frac{\mathbf{v}_{i,j}(x)^{2}\alpha _{j^{\prime }}(i)^{2}\cos\left\{t(\lambda _{i,j}+\widetilde{d}_{i}-\nu _{j^{\prime }})\right\}}{\sum _{l=1}^{k}n_{l}\alpha _{j^{\prime }}(l)^{2}}
&\leq
2\left(1-\displaystyle\sum _{j=1}^{k}\frac{\alpha _{j}(i)^{2}}{\sum _{l=1}^{k}n_{l}\alpha _{j}(l)^{2}}\right)\displaystyle\sum _{j^{\prime }=1}^{k}\frac{\alpha _{j^{\prime }}(i)^{2}}{\sum _{l=1}^{k}n_{l}\alpha _{j^{\prime }}(l)^{2}}\notag\\
&=
\frac{2}{n_{i}}\left(1-\frac{1}{n_{i}}\right).
\end{align*}
As a consequence, we have the following estimation on the return probabilities:
\begin{align*}
\left|P_{G_{n,t}}^{x}(x)-P_{G_{n_{i}},t}^{x}(x)\right|\leq \frac{4}{n_{i}},
\end{align*}
for $x\in V(G_{n_{i}})$. This implied that 
\begin{align*}
\lim_{n_{i}\to \infty }P_{G_{n,t}}^{x}(x)=\lim_{n_{i}\to \infty }P_{G_{n_{i}},t}^{x}(x),
\end{align*}
if $\lim_{n_{i}\to \infty }P_{G_{n_{i}},t}^{x}(x)$ exists. 
\\
\qed

\section{Local subgraph structure can cause localization}

In this section, we show two examples of CTQWs which cause strong localization for some vertices.

\subsection{Graphs with dominating vertices}

In this paper, we call a vertex $i\in V(G_{n})$ ``dominating vertex'' if $d(i)=n-1$, i.e., the vertex is connected with all other vertices in $V(G_{n})$. If there are $n_{d}$ numbers of dominating vertices in $G_{n}$, then the dominating vertices form a complete graph $K_{n_{d}}$ on $n_{d}$ numbers of vertices as an induced subgraph of $G_{n}$ (In other words, the induced subgraph of all dominating vertices is a clique $K_{n_{d}}$). In this case, $G_{n}$ is devided into two subgraphs $K_{n_{d}}$ and $G_{n-n_{d}}$ the induced subgraph with the vertex set $V(G_{n})\setminus V(K_{n_{d}})$. It is easy to see that $(K_{n_{d}},\ G_{n-n_{d}})$ is a fully interconnected graph decomposition of $G_{n}$. Therefore, we can apply Lemma \ref{lem:transition} with $k=2,\ G_{n_{1}}=K_{n_{d}},\ G_{n_{2}}=G_{n-n_{d}},\  \widetilde{d}_{1}=n-n_{d},\  \widetilde{d}_{2}=n_{d}$. 

The eigenvalues $\{\lambda _{1,l_{1}}\}_{l_{1}=1,\ldots ,n_{d}}$ and corresponding orthonormal eigenvectors $\{\mathbf{v}_{1,l_{1}}\}_{l_{1}=1,\ldots ,n_{d}}$ of $L_{K_{n_{d}}}$ are know as follows:
\begin{align*}
&\lambda_{1,l_{1}}=n_{d},
\quad 
\mathbf{v}_{1,l_{1}}=
\frac{1}{\sqrt{l_{1}(l_{1}+1)}}
\begin{bmatrix}
\mathbf{1}_{l_{1}}\\
-l_{1}\\
\mathbf{0}_{n-l_{1}-1}
\end{bmatrix}
\quad (l_{1}=1,\ldots ,n_{d}-1),\\
&\lambda_{1,n_{d}}=0,
\quad 
\mathbf{v}_{1,n_{d}}=
\frac{1}{\sqrt{n_{d}}}\ \mathbf{1}_{n_{d}},
\end{align*}
where $\mathbf{0}_{n}$ is the $n$ dimensional all zero vector. On the other hand, it is easy to see that the eigenvalues $\nu _{1}, \nu _{2}$ and corresponding eigenvectors $\mathbf{x}_{1}, \mathbf{x}_{2}$ are given as follows:
\begin{align*}
&\nu _{1}=n,
\quad 
\mathbf{x}_{1}=
\begin{bmatrix}
(n-n_{d})\mathbf{1}_{n_{d}}\\
-n_{d}\mathbf{1}_{n-n_{d}}
\end{bmatrix}
\\
&\nu _{2}=0,
\quad 
\mathbf{x}_{2}=\mathbf{1}_{n}.
\end{align*}
This shows that $\alpha _{1}(1)=n-n_{d},\  \alpha _{1}(2)=-n_{d},\  \alpha _{2}(1)=\alpha _{2}(2)=1$. Therefore from Eq.\ (\ref{expressU2}), we have the following result:
\begin{pro}[Dominating vertices can cause strong localization]\label{pro:dominating}
Let $G_{n}$ be a graph with arbitrary numbers of dominating vertices. If we consider CTQW starting from a dominating vertex $x$ then 
\begin{align*}
P_{G_{n},t}^{x}(y)
=
\begin{cases}
\displaystyle1-\frac{2}{n}\left(1-\frac{1}{n}\right)(1-\cos nt)
&\text{if $x=y$},
\\
\displaystyle\frac{2}{n^{2}}(1-\cos nt)
&\text{if $x\neq y$}.
\end{cases}
\end{align*}
Therefore 
\begin{align*}
\lim_{n\to \infty }P_{G_{n},t}^{x}(x)=1. 
\end{align*}
\end{pro}
\begin{rem}
Proposition \ref{pro:dominating} shows that if we consider the CTQW defined by the Laplacian matrix on complete graph then strong localization always occur for all vertices. Because complete graphs, star graphs and the threshold network model have dominating vertices, then CTQWs starting from dominating vertices on these graphs have the same transition probabilities.
\end{rem}

\subsection{Graphs with growing clique}
In this subsection, we show another sufficient condition for strong localization. Suppose $G_{n}$ includes a clique $K_{n_{c}}$. A vertex $v\in K_{n_{c}}$ is said to be a gateway vertex when there exist at least one edge $(v,w)\in E(G_{n})$ with $w\in V(G_{n})\setminus V(K_{n_{c}})$. 
\begin{pro}[Clique can cause strong localization]\label{pro:clique}
Suppose $G_{n}$ includes a clique $K_{n_{c}}$. Let $n_{g}$ be the number of gateway vertices in $K_{n_{c}}$ and $\{i_{1}, \ldots ,i_{n_{g}}\}\subset K_{n_{c}}$ be the set of all gateway vertices in $K_{n_{c}}$. If $x\in V(K_{n_{c}})\setminus \{i_{1}, \ldots ,i_{n_{g}}\}$ and $(n_{c}-n_{g})\to \infty $, then 
\begin{align*}
\lim_{n\to \infty }P_{G_{n},t}^{x}(x)=1. 
\end{align*}
\end{pro}
(Proof of Proposition \ref{pro:clique})\\
By the assumption, $(K_{n_{c}-n_{g}}, \{i_{1}\}, \ldots , \{i_{n_{g}}\}, G_{n-n_{c}})$ is a fully interconnected graph decomposition of the graph $G_{n}$, where $K_{n_{c}-n_{g}}$ be the clique with the vertex set $V(K_{n_{c}})\setminus \{i_{1}, \ldots ,i_{n_{g}}\}$ and $G_{n-n_{c}}$ be the induced subgraph with the vertex set $V(G_{n})\setminus V(K_{n_{c}})$. By Proposition \ref{pro:dominating}, we can see that 
\begin{align*}
\lim_{(n_{c}-n_{g})\to \infty }P_{K_{n_{c}-n_{g}},t}^{x}(x)=1. 
\end{align*}
Therefore, by the virtue of Theorem \ref{thm:localization}, we have desired result. 
\qed

\par
\
\par\noindent
{\bf Acknowledgments.} 
This work was supported by the Grant-in-Aid for Young Scientists (B) of Japan Society for the Promotion of Science (Grant No. 23740093). The author thank professors Norio Konno, Iwao Sato and Etsuo Segawa for fruitful discussions on this topic.


\begin{small}

\end{small}

\end{document}